# A Critical Review of Cyber-Physical Security for Building Automation Systems

Guowen Li[a], Lingyu Ren[b], Yangyang Fu[a], Zhiyao Yang[a], Veronica Adetola[c], Jin Wen[d], Qi Zhu[e], Teresa Wu[f,g], K. Selcuk Candan[f,h], Zheng O'Neill[a,*]

[a] *J. Mike Walker'66 Department of Mechanical Engineering, Texas A&M University, College Station, TX, USA*

[b] *Raytheon Technologies Research Center, East Hartford, CT, USA*

[c] *Pacific Northwest National Laboratory, Richland, WA, USA*

[d] *Department of Civil, Architectural, and Environmental Engineering, Drexel University, Philadelphia, PA, USA*

[e] *Department of Electrical and Computer Engineering, Northwestern University, Evanston, IL, USA*

[f] *School of Computing and Augmented Intelligence, Arizona State University, AZ, USA*

[g] *ASU-Mayo Center for Innovative, Arizona State University, AZ, USA*

[h] *Center for Assured and Scalable Data Engineering, Arizona State University, AZ, USA*

## Abstract

Modern Building Automation Systems (BASs), as the brain that enable the smartness of a smart building, often require increased connectivity both among system components as well as with outside entities, such as the cloud, to enable low-cost remote management, optimized automation via outsourced cloud analytics, and increased building-grid integrations. As smart buildings move towards open communication technologies, providing access to BASs through the building's intranet, or even remotely through the Internet, has become a common practice. However, increased connectivity and accessibility come with increased cyber security threats. BASs were historically developed as closed environments with limited cyber-security considerations. As a result, BASs in many buildings are vulnerable to cyber-attacks that may cause adverse consequences, such as occupant discomfort, excessive energy usage, and unexpected equipment downtime. Therefore, there is a strong need to advance the state-of-the-art in cyber-physical security for BASs and provide practical solutions for attack mitigation in buildings. However, an inclusive and systematic review of BAS vulnerabilities, potential cyber-attacks with impact assessment, detection & defense approaches, and cyber resilient control strategies is currently lacking in the literature. This review paper fills the gap by providing a comprehensive up-to-date review of cyber-physical security for BASs at three levels in commercial buildings: management level, automation level, and field level. The general BASs vulnerabilities and protocol-specific vulnerabilities for the four dominant BAS protocols (i.e., BACnet, KNX, LonWorks, and Modbus) are reviewed, followed by a discussion on four attack targets and seven potential attack scenarios. The impact of cyber-attacks on BASs is summarized as signal corruption, signal delaying, and signal blocking. The typical cyber-attack detection and defense approaches are identified at the three levels. Cyber resilient control strategies for BASs under attack are categorized into passive and active resilient control schemes. Open challenges and future opportunities are finally discussed.

| **Nomenclature** | | | |
|---|---|---|---|
| AEAD | Authenticated Encryption with Associated Data | LAN | Local Area Network |
| AHU | Air Handling Unit | IDS | Intrusion Detection System |
| ANN | Artificial Neural Networks | IoT | Internet of Things |
| ASHRAE | American Society of Heating, Refrigerating and Air Conditioning Engineers | IP | Internet Protocol |
| ATT&CK | Adversarial Tactics, Techniques, and Common Knowledge | IPSec | Internet Protocol Security |
| BACnet | Building Automation and Control Networking Protocol | IT | Information Technology |
| BACnet/SC | BACnet Secure Connect | MPC | Model Predictive Control |
| BASs | Building Automation Systems | MIMO | Multiple-Input–Multiple-Output |
| BMSs | Building Management Systems | MITM | Man-In-The-Middle |
| CPSs | Cyber-Physical Systems | OSI | Open Systems Interconnection |
| CTD | Cyber Threat Dictionary | OT | Operational Technology |
| DoS | Denial of Service | SCADA | Supervisory Control And Data Acquisition |
| DDoS | Distributed Denial of Service | SMPC | Stochastic Model Predictive Control |
| FDD | Fault Detection and Diagnosis | SQL | Structured Query Language |
| FTCS | Fault-Tolerant Control System | SSL/TLS | Secure Sockets Layer and Transport Layer Security |
| GAN | Generative Adversarial Networks | SSS | Sub-keyword Synonym Searching |
| GEBs | Grid-interactive Efficient Buildings | TCP | Transmission Control Protocol |
| HIL | Hardware-In-the-Loop | NIST | National Institute of Standards and Technology |
| HVAC | Heating, Ventilation, and Air Conditioning | OSI | Open Systems Interconnection |
| ISP | Internet Service Provider | VPN | Virtual Private Network |
| ISRA | Information Security Risk Analysis | WAN | Wide Area Network |
| KPI | Key Performance Index | XSS | Cross-Site Scripting |

# 1. Introduction

According to the Intelligent Building Institute of the United States, an Intelligent Building (or Smart Building) is one that "provides a productive and cost-effective environment through optimization of its four basic elements including structures, systems, services and management and the interrelationships between them (Wigginton & Harris, 2013)." Building Automation System (BAS) serves as the brain for intelligent buildings. It includes cyber-infrastructure components of sensing, computation, communication, and control that provide close monitoring and operations for the mechanical and energy systems, and physical environment in buildings. A BAS is defined as "an automated system where building services, such as utilities, communicate with each other to exchange digital, analog or other forms of information, potentially to a central control point (Brooks, Coole, Haskell-Dowland, Griffiths, & Lockhart, 2017)."

With the increasing usage of remote/mobile access, integrated wearable technologies, data exchange, and cloud-based data analytics in modern intelligent buildings, the BAS moves towards open communication technologies. Providing access to the BAS through the building's intranet, or even remotely through the Internet, has become a common practice.

BASs were historically developed as closed environments. BACnet (Liaisons, et al., 2012), the most popular communication protocol for BAS in commercial buildings, was not designed with security as a primary requirement because: (1) the original intention and implementation of BASs were isolated from external connections (Peacock, 2019); and (2) physical wiring was typically installed without easily accessible sockets as we find today with Ethernet installations. Hence, security did not play a particular role in the original design of BAS. Today, it is challenging to enhance the legacy BAS protocols with appropriate mechanisms because the existing BAS architecture does not provide sufficient hardware and software resources for these adaptations. For example, a challenging problem for implementing security approaches is the limitation of BAS field devices. Even when existing standards allow for extensions, full-blown security mechanisms need computing resources and time for execution, which are typically unavailable on field devices (Sauter, Soucek, Kastner, & Dietrich, 2011).

Since the originally isolated BASs were designed with little cyber-security considerations, BASs could be attack targets. Several known real-world cyber-attacks (Griffiths, 2014, Higgins, 2021, Koh, 2018, Kumar, 2016, McMullen, Sanchez, & Reilly-Allen, 2016, Molina, 2015, Zetter, 2013) on buildings were reported from 2013 to 2021, as shown in Figure 1. In May 2013, the BAS of Google Australia Office was hacked by two security researchers by exploiting BAS software vulnerabilities (Zetter, 2013). In November 2013, Target Corporation, a large retailer in the United States, saw its network hacked and broken into. The attacker utilized network credentials stolen from a vendor of refrigeration, heating and air conditioning equipment (McMullen, et al., 2016). In July 2014, the St. Regis Shenzhen 5-star hotel was hacked by a hacker who took control of around a hundred rooms in the hotel (Griffiths, 2014). The hotel's BAS had several flaws that allowed Molina (Molina, 2015) to create a remote control to access the hotel rooms. In October 2016, hackers used Distributed Denial of Service (DDoS) attack to shut down two apartments' heating systems in Finland (Kumar, 2016). In August 2018, a security engineer hacked into the WiFi of a hotel while attending a cybersecurity conference in Singapore. The engineer hacked into the server and blogged about it online, where he published the hotel administrator's server passwords (Koh, 2018). In December 2021, a firm located in Germany discovered that three-quarters of the BAS devices in the office building system network had been mysteriously locked down with the system's own digital security key, which was under the attackers' control. It suddenly lost contact with hundreds of its BAS devices including light switches, motion detectors, shutter controllers, etc. The firm had to revert to manually flipping on and off the central circuit breakers in order to power on the lights in the building (Higgins, 2021). As of 2019, 37.8% of computers used to control BASs were subject to some kind of malicious attacks according to Kaspersky's report (Kaspersky, 2019). The growing interest from adversary individuals and agents in BAS is driven by the deep integration of building services, especially the safety-critical (e.g., fire or social alarm systems) and security-critical (e.g., access control systems) services (Granzer, Praus, & Kastner, 2009). This integration enables low-cost functionality improvement via data sharing and cooperative control. However, it also breaks the physical isolation of the subsystems and thus enlarges the BAS cyber-attack surface (King, 2016). Furthermore, modern buildings are also capable of providing grid ancillary services, such as demand response and frequency regulation (Fu, O'Neill, Wen, Pertzborn, & Bushby, 2021). These buildings, also called Grid-interactive Efficient Buildings (GEBs), provide open doors to grid operations, which raise new security concerns. Therefore, there is a strong need to advance the state-of-the-art in cyber-physical security for intelligent buildings and provide solutions for attack mitigation.

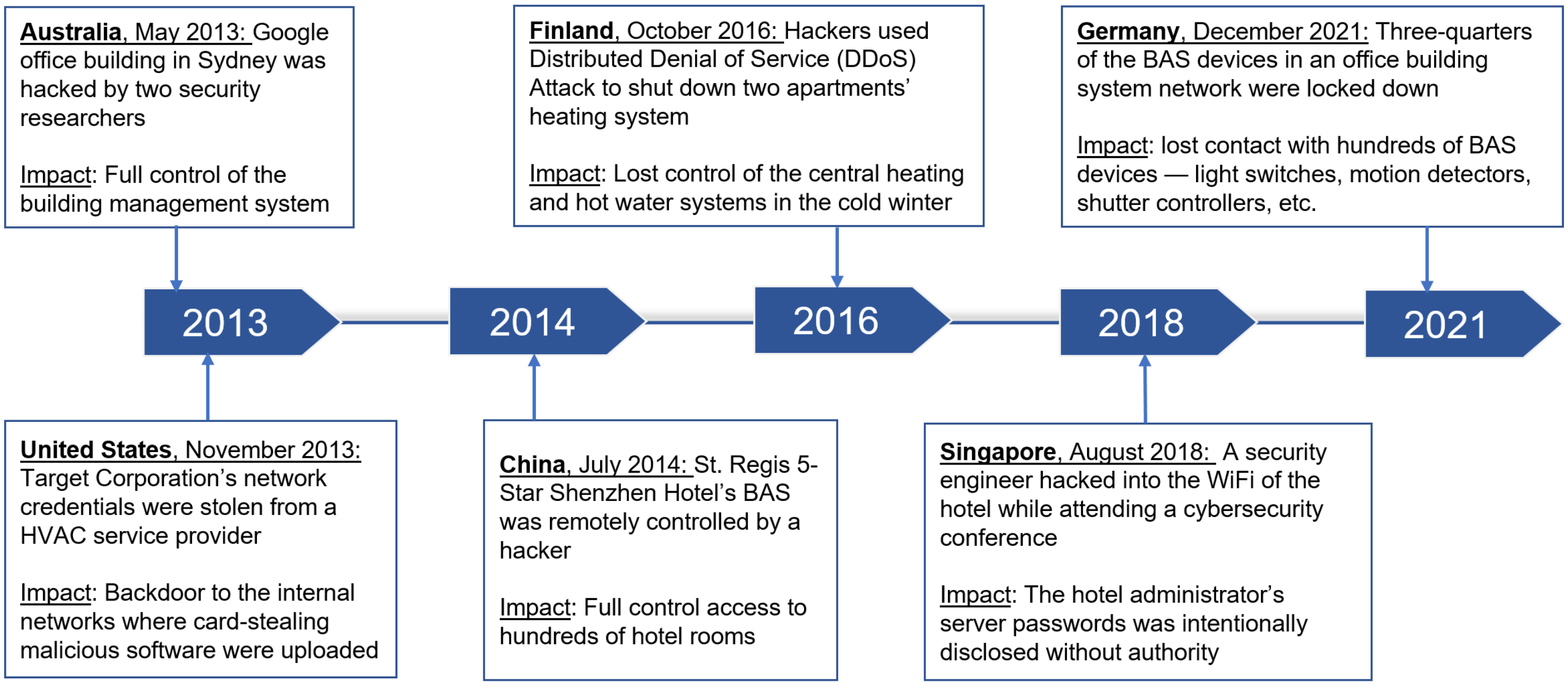

Figure 1. Timeline of recently reported cyberattacks on buildings and their physical impacts.

The International Telecommunications Union defines cyber security as "the collection of tools, policies, security concepts, security safeguards, guidelines, risk management approaches, actions, training, best practices, assurance and technologies that can be used to protect the cyber environment and organization and user's assets" (Von Solms & Van Niekerk, 2013). Cyber-physical security aims to address security concerns for physical systems including the Internet of Things (IoT), industrial control systems, and BASs. One early effort to establish BAS cyber security terminology defines two major classes of cyber-attacks based on the attack target: network attacks and device attacks (Granzer, Praus, et al., 2009). Network attacks refer to compromised access to either network medium or network devices, while device attacks refer to any direct physical or software attacks on edge devices. Subsequently, a three-level classification (management level, communication level, and automation level) model was presented in (Kharchenko, Ponochovnyi, Boyarchuk, & Qahtan, 2017) considering attacks and physical faults. Giraldo et al. (Giraldo, Sarkar, Cardenas, Maniatakos, & Kantarcioglu, 2017) also mentioned that the user privacy issue is one of the security concerns. For example, the SHODAN search engine (Matherly, 2015) can list BAS systems connected to the Internet, which could make them easy attack targets. Attackers can be motivated to attack a BAS so that they can gain access to the surveillance system (e.g., IP cameras) and thus violate user privacy. Qi et al. (Qi, Kim, Chen, Lu, & Wang, 2017) reviewed the cyber security challenges for the GEBs providing demand response services. The main concern is the potential physical influences on the power grid operation induced by malicious BAS control commands.

The rising demand for enhancing BAS cyber-security calls for a comprehensive understanding of the BAS cyber landscape. A few publications have been focused on cyber-physical security on BASs, which mainly cover cyber-attacks, detection, and defense related topics. dos Santos et al. (dos Santos, Dagrada, & Costante, 2021) demonstrated how to attack a BAS workstation via a smart lighting system and surveillance system, proving how deep integration increased the attack vectors. Wendzel et al. (Wendzel, Zwanger, Meier, & Szlósarczyk, 2014) presented a botnet scenario where compromised BAS devices are used as bots to allow massive aggregated attacks. Kaur et al. (Kaur, Tonejc, Wendzel, & Meier, 2015) focused on BACnet protocols and listed potential attacks in the BACnet network, such as network flooding, traffic redirection, and re-routing Denial-of-Service (DoS) attacks. Raiyn (Raiyn, 2014) discussed different types of cyber-attacks and listed typical attack detection strategies including intrusion detection systems (IDS), misuse detection, misbehavior detection, anomaly detection, and signature-based detection approaches. Yurekten and Demirci (Yurekten & Demirci, 2021) presented a systematic review of cyber threat categories and related defense approaches including defense against network scanning attacks, spoofing attacks, network-level DoS attacks, sniffer attacks, malware, and web application attacks. Ciholas et al. (Ciholas, Lennie, Sadigova, & Such, 2019) presented a systematic literature review of cyber-attacks, vulnerabilities, and defense approaches for smart buildings in terms of three levels (i.e., management, automation, and field levels), where common cyber-attacks (e.g., wireless attacks, DoS attacks, protocol-specific attacks, privacy attacks) and corresponding defense approaches were illustrated in detail. Graveto et al. (Graveto, Cruz, & Simöes, 2022) provided a systematic survey of the typical three-level BAS architecture with dominant protocols, BAS security risks with possible cyber-attacks, and proposals for BAS security enhancement including security monitoring, anomaly detection, IDS, etc. To maintain acceptable levels of system operation in the presence of cyber-attacks, the concept of cyber resilient control is proposed for cyber-physical systems. But few publications have focused on cyber resilient control strategies specifically for BASs in commercial buildings. Generally speaking, in contrast to other domains that recently received substantial attention such as industrial control and automation systems (Graveto, et al., 2022), the security of BASs has been discussed in a less structured manner. An in-depth analysis is still needed to systemically address the cyber-security issues of BASs in the context of the emerging openness and connectivity of intelligent buildings.

Although there are several reviews on cyber-physical security for BASs as mentioned above, to the authors' best knowledge, a holistic overview integrating BAS vulnerabilities, potential threats with impact assessment, cyber-attack detection & defense, and cyber resilient control is still missing in this field. To fill the research gap, this paper aims to provide insights into the following significant questions:

1. Why are BASs vulnerable to cyber-attacks?
2. What are the common cyber-attacks and their impact on BASs?
3. What are the existing approaches of cyber-attack detection and defense?
4. How do the existing cyber resilient control strategies work?
5. What are the research challenges and future opportunities?

The remainder of this paper is organized as shown in Figure 2. Section 2 introduces the literature review and evaluation method. Section 3 summarizes the literature review results of vulnerabilities, potential threats, detection & defense approaches, and resilient control strategies. Section 4 discusses the open challenges and future opportunities. Section 5 concludes this review work.

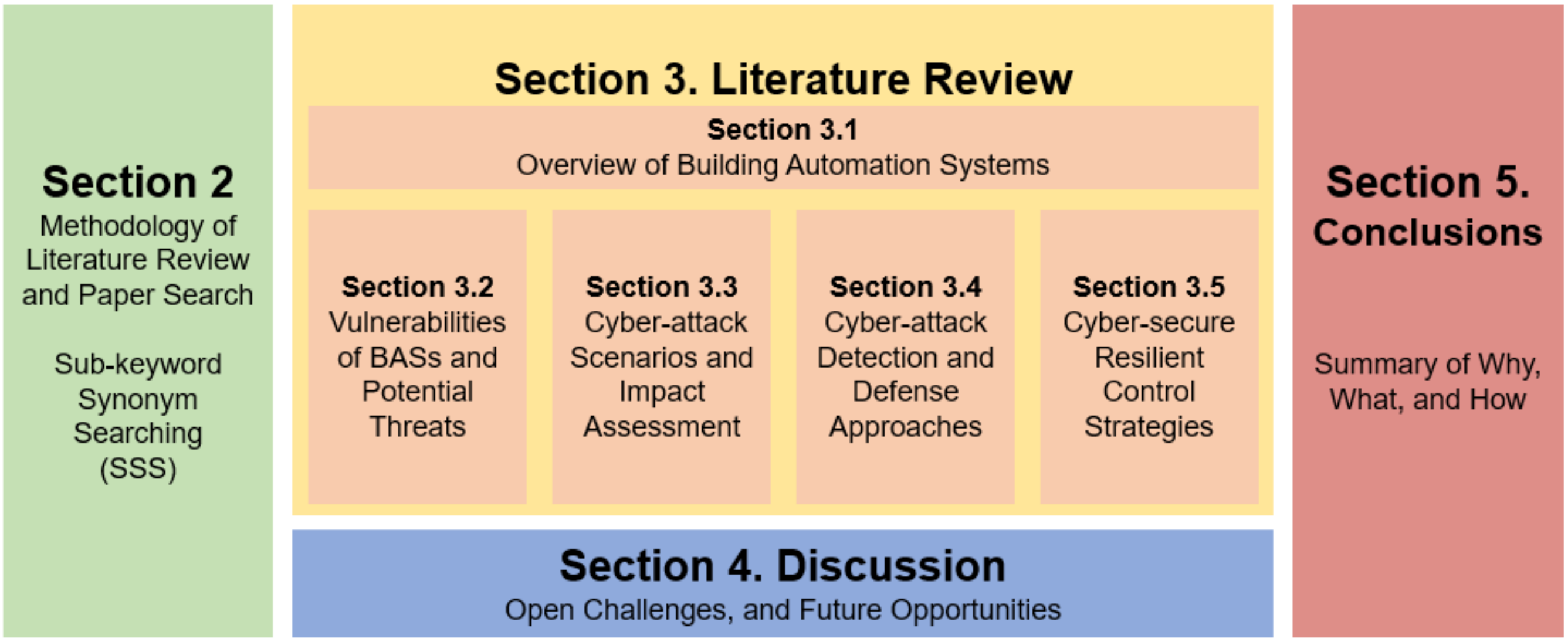

Figure 2. Content organization diagram of this review paper.

# 2. Methodology

## 2.1 Literature Review

To conduct a comprehensive review that captures the most important literature, we applied a searching methodology called Sub-keyword Synonym Searching (SSS) (Zhang, et al., 2021). In this paper, Google Scholar is the main search engine of the methodology, and the full list of searching keywords in Google Scholar is the full combination of each sub-keyword. The purpose of this methodology is to exhaustively identify relevant papers by multiple searches with synonym sub-keywords.

Table 1. Parameters of Sub-keyword Synonym Searching (SSS) in this review paper.

| Parameter | Values |
|---|---|
| Sub-keyword 1 | 'cyber security', 'cyber attack', 'attack detection', 'attack defense', 'securing', 'resilient control', 'control under attack' |
| Sub-keyword 2 | 'building automation system', 'building energy management', 'smart building', 'HVAC' |
| Number of papers per search | 20 |
| Year from | 2010 |
| Year to | 2022 |
| Citation threshold (2010-2021) | 3 |
| Citation threshold (2022-present) | 0 |

Table 1 summarizes the parameters of the SSS methodology used in this paper. The SSS methodology uses sub-keywords and synonyms to conduct multiple searches to comprehensively capture the most important papers in the same field. SSS makes sense because (1) different authors use different terms for the same concept and using synonyms can avoid missing papers with different terms, and (2) SSS can cover various sub-topics (e.g., cyber security, detection and defense, resilient control). The total searched papers are (7 × 4) keywords × (20) top papers found/keyword = 560 papers, and 302 is the final number after manually removing duplicates. The identified 302 papers with associated references were carefully reviewed, out of which over 110 papers were selected based on expert domain knowledge for this study. These selected papers are categorized and organized following the structure of this paper.

## 2.2 Review Statistics

Figure 3 (a) shows the word cloud of the reviewed literature titles. The terms, “cyber security”, “building”, “attack”, “detection”, “defense”, and “control” were among the most popular words from the reviewed articles. Figure 3 (b, c) shows the journal where the articles were published and the number of publications in recent years. In general, there is a growing trend of publications during 2010 - 2016. The 110 reviewed articles were published in 41 online resources, which mainly include Institute of Electrical and Electronics Engineers (IEEE) (46%), Association for Computing Machinery (ACM) (6%), Computer & Security (4%), International Journal of Critical Infrastructure Protection (IJCIP) (3%), Applied Energy (2%), International Federation of Automatic Control (IFAC) (2%) and several reports from American Society of Heating, Refrigerating and Air-Conditioning Engineers (ASHRAE) (4%) and National Institute of Standards and Technology (NIST) (2%). The major topics’ distribution is BAS Vulnerabilities related topic (14%), Potential Threats and Impact Assessment related topic (15%), Detection related topic (24%), Defense related topic (14%), and Resilient Control related topic (13%).

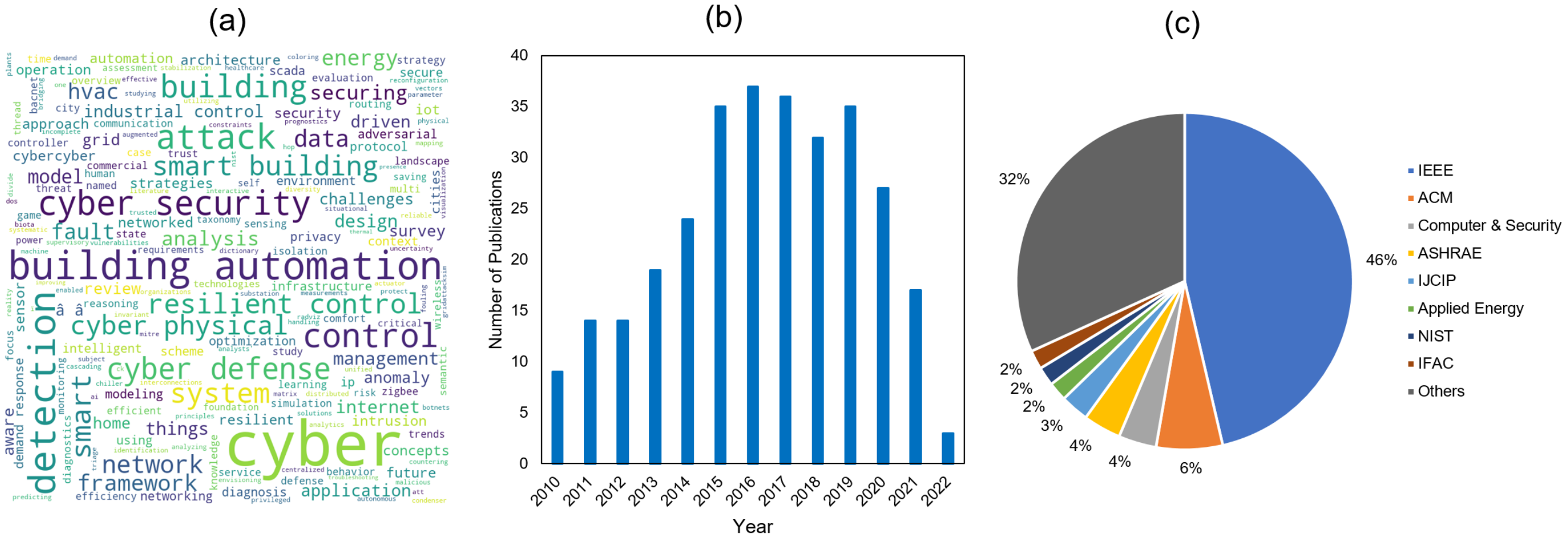

Figure 3. Overview of the reviewed literature: (a) word cloud of the selected literature titles, (b) publications by years, and (c) publications by sources.

# 3. Results of the Review

Section 3 focuses on the current state-of-the-art in five aspects, 1) overview of building automation systems, 2) vulnerabilities of BAS and potential threats, 3) cyber-attack scenarios in BASs and impact assessment, 4) cyber-attack detection and defense approaches, 5) cyber resilient control for BASs.

## 3.1 Overview of Building Automation Systems

BAS is in charge of the automatic control of a building's heating, ventilation, and air conditioning (HVAC) and other systems including the security, fire safety, and lighting systems. Some major objectives of BAS are to maintain occupant comfort, increase building energy efficiency, reduce building energy consumption, enhance demand flexibility, and prolong the life span of building equipment (Salsbury, 2005). As the number of devices in a building grows, BAS vendors integrate Internet Protocol (IP) and open standards, such as BACnet (Building Automation and Control Networking Protocol), to manage the network of devices (Newman, 2013). The European Committee for Standardization divides building automation architecture and communications into three levels: Management Level, Automation Level, and Field Level (EN/ISO, 2017).

- The *Management Level* represents the information technology and communication network. This level comprises operator stations, monitoring and operator units, programming units, and other peripheral computer devices connected to a data processing device (i.e., a server) to support the information exchange monitoring and management of the automation system. In general, the Management level contains the human interface (e.g., workstations), server, and routing devices, all connected via an appropriate communication medium, such as LAN/WAN (Local Area Network/Wide Area Network) using TCP/IP (Transmission Control Protocol/Internet Protocol) or BACnet/IP (Brooks, et al., 2017).
- The *Automation Level* corresponds to a dedicated communication network for the sole purpose of building device connectivity, communication, and control. This level is associated with controllers that serve main plants, such as air handling units, chillers, boiler units, etc. The Automation level provides the various primary control technology devices and secondary facility automation, connected via networked controllers and operating via communication protocols, such as BACnet, LonWorks (Loy, Dietrich, & Schweinzer, 2001), or KNX (Konnex) (Ruta, Scioscia, Loseto, & Di Sciascio, 2017).
- The *Field Level* includes sensors, activators, and devices connected to the specific plant and equipment. These devices are generally self-contained physical units, either application-specific or generic controllers. Application-specific controllers' operation uses communication protocols, such as Modbus (Thomas, 2008), KNX, or other proprietary protocols.

Figure 4 illustrates the BAS architecture in terms of these three levels. An advantage of such an architecture is a clear separation of duties and a reduction of network traffic at the management level. However, for smaller systems, the separation of networks can be expensive (Brooks, et al., 2017).

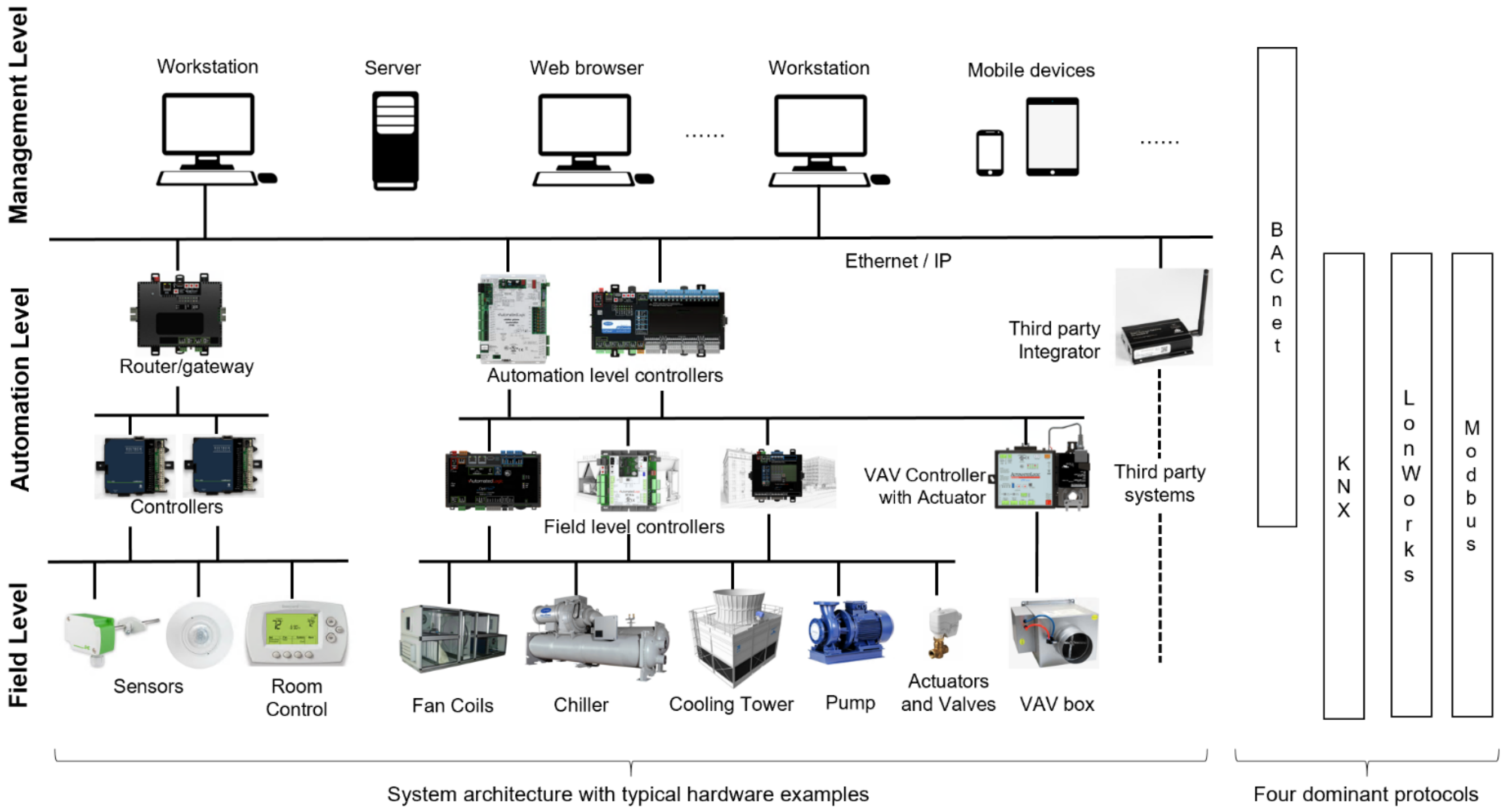

Figure 4. Three-level BAS architecture and the dominant protocols for each level (adapted from (Brooks, et al., 2017, Merz, Hansemann, & Hübner, 2009)).

### 3.2 Vulnerabilities of BASs and Potential Threats

Section 3.2 reviews the vulnerabilities and known threats for BASs and the protocol-specific vulnerabilities. Prior efforts have defined and used different taxonomies to classify the vulnerabilities and threats in BASs. For example, early research (Granzer, Praus, & Kastner, 2010) used attack targets to categorize attacks into network attacks and device attacks. This taxonomy is extended in a recent work (Liu, Pang, Dán, Lan, & Gong, 2018) where five phases of the security interaction between devices and building automation network were added to further explain the security requirements in the life cycle of a BAS device. Anwar et al. (Anwar, Nazir, & Mustafa, 2017) used a simple taxonomy that groups cyber-attacks into unintentional, international, and malfunction. Mundt and Wickboldt (Mundt & Wickboldt, 2016) summarized the security findings of BASs in three levels (i.e., management level, automation level, and field level). In this research, based on the network and physical features, we elaborate on how each level of the BAS can be vulnerable to different attacks. The details of the attacks are shown in Table 2.

Table 2. List of common cyber-attacks on BASs (Faraji Daneshgar & Abbaspour, 2016, Gupta & Gupta, 2017, Pan, Pacheco, & Hariri, 2016, Pingle, Mairaj, & Javaid, 2018, Rohatgi, 2009).

| Attack Type | Attack Description |
|---|---|
| Cross-Site Scripting (XSS) attack | This is a type of injection, in which malicious scripts are injected into otherwise benign and trusted websites. The malicious script can access cookies, session tokens, or other sensitive information retained by the browser and used with that site. |
| Denial-of-Service (DoS) attack / Distributed Denial-of-Service (DDoS) attack | A DoS attack means to shut down a machine or network, making it inaccessible to its intended users, by flooding the target with traffic or sending it information that triggers a crash. A DDoS attack is a malicious attempt to disrupt the normal traffic of a targeted server, service, or network by utilizing multiple compromised computer systems as sources of attack traffic. |
| Electromagnetic attack | This is a side-channel attack performed by measuring the electromagnetic radiation emitted from a device and performing signal analysis on it. |
| Fuzzing attack | This is an automated process used to find application vulnerabilities. It consists of inserting massive amounts of random data into source code and observing the outcomes. |
| Man-In-The-Middle (MITM) attack | MITM is a type of attack in which a third party in stealth takes control of the communication channel between two or more parties. In MITM attack, the attacker can intercept, modify, change, or replace target victim's communication traffic while the victims are not aware of the man in the middle. |
| Password Brute-Force attack | This is an attempt to discover a password by systematically trying every possible combination of letters, numbers, and symbols until you discover the one correct combination that works. |
| Replay attack | This is a form of network attack in which valid data transmission is maliciously or fraudulently repeated or delayed. This is carried out either by the originator or by an adversary who intercepts the data and re-transmits it, possibly as part of a spoofing attack by IP packet substitution. |
| Sniffing attack | Sniffing corresponds to the theft or interception of data by capturing the network traffic using a packet sniffer (an application aimed at capturing network packets). |
| Snooping attack | This type of attack can involve an intruder listening to the network traffic. If traffic includes passing unencrypted passwords, an unauthorized individual can potentially access the network and read confidential data. |
| Spoofing attack | Spoofing is a situation in which a person or program successfully identifies as another by falsifying data, to gain an illegitimate advantage. |
| Structured Query Language (SQL) injection attack | This attack uses malicious SQL code for backend database manipulation to access information that is not intended to be displayed. It can read sensitive data from the database, modify database data, and execute administration operations on the database. |

### 3.2.1 BAS Vulnerabilities and Threats

***Management Level***

The network and devices at the management level are often Information Technology (IT) based systems and are vulnerable to known IT threats (Ciholas, et al., 2019): web-based building management systems are vulnerable to Structured Query Language (SQL) injection attacks, password attacks, cross-site scripting, or DoS attacks if not configured properly. A workstation with email services may be exposed to phishing attacks where malicious codes (such as Trojans (Xiao, et al., 2016)) can be delivered and planted as backdoor malware. The credentials of the management software are often shared among vendors, clients, and field engineers for installation and maintenance. This access control and authorization policy opens a gate for low-effort insider attacks. Once the attackers gain access to the management devices, they are empowered with supervisory-level controls and can potentially damage the whole BAS.

***Automation Level***

The automation level controllers are vulnerable to both remote attacks and local attacks (Brooks, et al., 2017). The remote attackers can leverage the covert channels on the management devices to inject malware on the controller or maliciously reprogram the control logic. Meanwhile, the attackers could also perform DDoS from the local botnet devices. Unlike IT-based networks, automation-level network devices are less equipped with state-of-the-art intrusion detection systems or firewalls. Moreover, the current implementations of BAS protocols lack basic authentication and encryption, which makes it possible to perform snooping attacks, network rerouting attacks, malicious data injection, and replay attacks (Holmberg & Evans, 2003). These protocol-specific threats are reviewed in Section 3.2.2 Protocol-specific Vulnerabilities and Threats.

***Field Level***

Without a strong physical access control policy, field-level devices are more exposed to near-field attacks. For example, electromagnetic side channel attacks can monitor electromagnetic emissions and reverse engineer the signals for information leakage (Rohatgi, 2009). If the devices are wirelessly connected, they may be the target of man-in-the-middle attacks that hijack the wireless channel from the router. Most field devices are embedded systems that are vulnerable to hardware/firmware attacks. For example, one could connect with the serial port of a sensor and change the firmware configurations to generate incorrect sensing data. Due to limited computing and memory resources, these devices are also vulnerable to continuous fuzzing attacks which may drain their battery or crash their processors. Mundt and Wickboldt (Mundt & Wickboldt, 2016) provided a detailed security inspection of a real-world BAS system. Specifically, for the field level, they found a few attack vectors: (1) there are open LON (Local Operating Network) interfaces for covert device connections, and the network is not zoned for different levels of authorization; (2) the electromagnetic emission of the KNX signals on twisted pair cables can be captured by a simple antenna and decoded by audio equipment, which leads to data leakage; (3) when correlating the physical actions (e.g., switch On/Off lights) with the detected signals, it is possible to discover device addresses and positions.

### 3.2.2 Protocol-specific Vulnerabilities and Threats

This section further reviews vulnerabilities and threats that are specific to four dominant BAS protocols: BACnet, KNX, LonWorks, and Modbus.

***BACnet (Management & Automation Level)***

BACnet is an open protocol developed by ASHRAE. BACnet is designed with Internet connection capability, thus BACnet networks can be exposed to remote attackers. The generic protocol design vulnerabilities of BACnet were discussed in (Holmberg & Evans, 2003, Kaur, et al., 2015). These vulnerabilities are mostly caused by the lack of authentication and encryption. Potential threats include snooping attacks that eavesdrop on network identity or device property information, network rerouting, network or application layer DoS attacks, and direct application service attacks that inject erroneous data into the system.

***KNX (Automation & Field Level)***
KNX is a standardized OSI (Open Systems Interconnection) based protocol that allows different physical transmission mediums. One known issue is that KNX transmits passwords using plaintext, which was exploited in (Antonini, Barenghi, Pelosi, & Zonouz, 2014) for password sniffing attacks. KNXnet/IP is a version of KNX that encapsulates the payload in IP stack, which makes it possible for KNX devices to report to management devices through Ethernet connection. As the original KNX is designed for local networks with little security consideration, the KNXnet/IP relies heavily on the IP network security measures, such as IPSec (Internet Protocol Security), SSL/TLS (Secure Sockets Layer and Transport Layer Security), and VPN (Virtual Private Network). None of these security solutions can fully protect the communication within a KNXnet/IP network, and the researchers in (Lechner, Granzer, & Kastner, 2008) proposed a new security extension located between the automation level and the field level to provide authenticated and encrypted communication channels.

***LonWorks (Automation & Field Level)***
LonWorks (or Local Operating Network) is an open standard (ISO/IEC 14908) designed for building automation systems. LonWorks network supports a single shared key among all devices and employs a challenge-response protocol to ascertain if a device is part of the network. The application data is not encrypted nor provided with any integrity checks (Antonini, Maggi, & Zanero, 2014). Due to the weak password policy and non-protected payload, LonWorks is generally vulnerable to password brute-force attacks, DDoS, information disclosure, and spoofing attacks (Kamal, Abuhussein, & Shiva, 2017).

***Modbus (Automation & Field Level)***
Modbus is a serial communication protocol commonly used in industrial control systems. The Modbus serial driver is vulnerable to stack-based buffer overflow attacks as reported in ((ICSA-14-086-01A), 2018). When used as the application layer of a TCP/IP stack on Ethernet, Modbus is not protected by any cryptographic primitive (Antonini, Barenghi, et al., 2014). Chen et al. (Chen, Pattanaik, Goulart, Butler-Purry, & Kundur, 2015) performed DoS attacks using TCP SYN Flood and man-in-the-middle (MITM) attacks using Ettercap for Modbus/TCP implemented in a lab testbed.

In summary, all major BAS protocols lack strong authentication and encryption mechanisms in their design, which makes them vulnerable to various versions of service accessibility attacks and data confidentiality & integrity attacks.

## 3.3 Cyber-attack Scenarios in BASs and Impact Assessment

### 3.3.1 Attack Scenarios

This section introduces the attack targets under a BAS IT/OT (Information Technology/Operational Technology) framework and defines attack scenarios for the cyber-physical security of BASs. Sensors, actuators, and controllers in the BAS are connected through the OT network while management workstations and servers are connected through the IT network. A majority of the devices on the OT network are exposed to the Internet through IT connections. However, some subsystems could have direct access to the Internet to allow remote vendor support. Either of these connections can be leveraged by remote hackers to penetrate the target BAS, as shown as purple dashed lines in Figure 5. Overall, these attacks could target four components (Granzer, Praus, et al., 2009):

- *Target 1: Management devices running on IT network.* An adversary could target the servers and workstations where major functions, such as monitoring, scheduling, energy saving, and event responding, are performed and subsystems are integrated and synergized.
- *Target 2: Interface from IT to OT network.* An adversary on the IT network may hijack the legal IT-to-OT conversation via MITM attacks or false data injections (pretending to be the server). The attacker can then perform eavesdropping or malicious device controls.
- *Target 3: Interface from OT to IT network.* Similar to the previous attack target, an adversary on the OT network could target the OT-to-IT interface by stealing the device ID and pretending to be one of them.
- *Target 4: Field devices running on OT network.* An adversary could also target field devices to damage the device, interrupt building operations, or even impact power grids through aggregated building device controls.

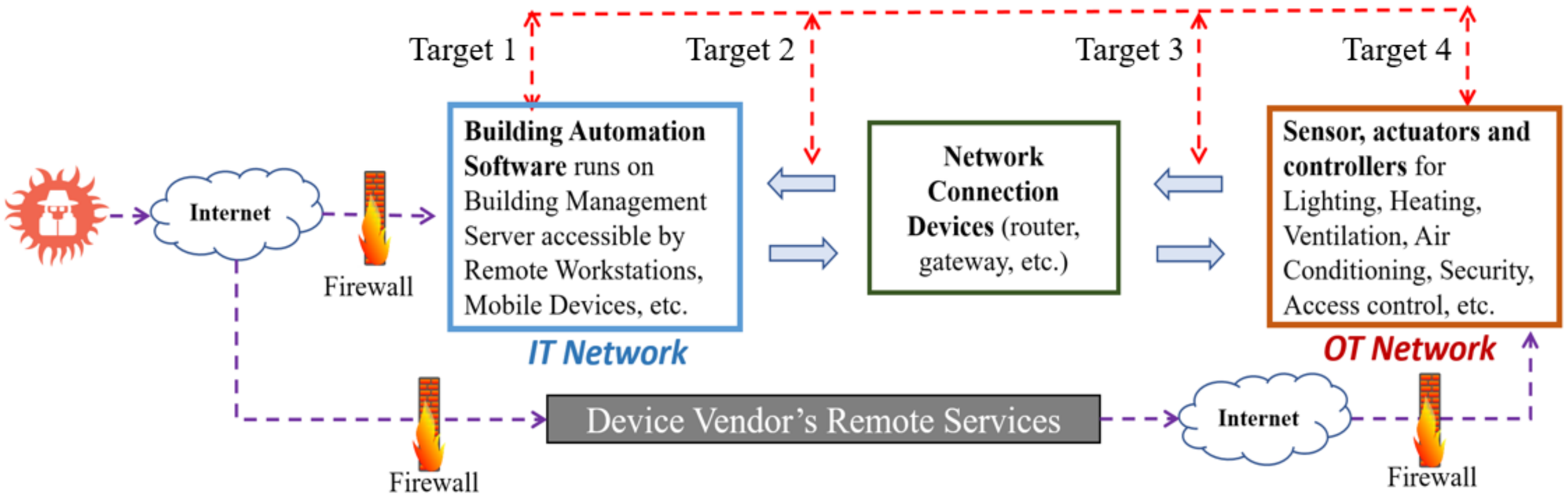

Figure 5. Potential attack targets in BASs.

Based on whether the attacks interrupt the network communication, they can be classified into two major categories: passive attacks that try to obtain data exchanged in the network without interrupting the communication, and active attacks that lead to the disruption of the normal functionality of the network, usually with information interruption, modification, or fabrication. Examples of passive attacks include eavesdropping, traffic analysis, and traffic monitoring. Examples of active attacks include jamming, impersonating, modification, DoS, and message replay (Abdel-Fattah, Farhan, Al-Tarawneh, & AlTamimi, 2019). Based on the attack targets, we define a list of attack scenarios in Table 3 that are most common and impactful to BASs. The attack implementations are given using BACnet as the example, but they can be extended to other protocols.  These scenarios can be grouped into four categories:

- *Reconnaissance Attacks (scenarios 1 and 2)* where attackers gather information and identify attack vectors.
- *Availability Attacks (scenarios 3, 4, 5)* where attackers partially or fully disable the target device from its regular tasks.

- *Covert Channel Attacks (scenario 6)* where attackers plant malware on the device and create covert channels to allow long-term persistent attacks.
- *Function Attacks (scenario 7)* where attackers deliver malicious payloads.

Table 3. Typical scenarios of BAS attacks (Holmberg & Evans, 2003, Kaur, et al., 2015).

| Scenario | Attack Description | Implementation (BACnet) | Attack Type | Impact | Attack Level |
|---|---|---|---|---|---|
| 1 | Network Mapping | Sending probes (Who-Is, Who-Is-Router) | passive | information exposure | Automation & Field Levels |
| 2 | Device Fingerprinting | Sending ReadProperty message to gain information about the device | passive | information exposure | Automation & Field Levels |
| 3 | Network DoS | (1) Modify SADR (source address) field and craft unknown message type so that the router answers Reject-Message-To-Network to the broadcast address; (2) traffic redirection to a target router; (3) use Router-Busy-To-Network message to spoofed router; (4) use Initialize-Routing-Table message to create a dead loop between routers; (5) send I-Am-Router-To-Network message to redirect traffic; (6) send Initialize-Routing-Table message | active | lose availability of network routers or network links | Automation & Field Levels |
| 4 | Device DoS | (1) Use the I-Am service to pretend to be another device; (2) use Who-Is to flood the network so that all devices busy answering I-Am; (3) use re-initialize to reboot unsecure devices; (4) traffic redirection to a target device | active | lose availability of target device | Automation & Field Levels |
| 5 | Server DoS | (1) Flood the web server with requests from edge devices; (2) software attack (malformed payload to create buffer overflow) | active | lose availability of central controller or web server | Management Level |
| 6 | Device Backdoor | Hide malicious commands in payload and use WriteProperty to communicate with a remote attacker | active | allow persistent remote access and control to target devices | Automation & Field Levels |
| 7 | Remote-to-Device | Use WriteProperty to change control settings or turn on/off devices | active | allow physical control | Field Level |

### 3.3.2 Impact Assessment of Cyber-attacks on BASs

The impacts of cyber-attacks on BASs can be summarized as signal corruption, signal delaying, and signal blocking.

- *Signal Corruption* refers to the manipulation of communicated data through remote attacks that can utilize services like WriteProerty to corrupt the value of the payloads. Huang et al. (Huang, et al., 2009) provided basic models for signal corruptions, such as max/min attack, scaling attack, and additive attack. Sridhar and Manimaran (Sridhar & Govindarasu, 2014) extended the basic signal corruption patterns to include ramp attack, pulse attack, and random attack.
- *Signal Delaying*, which is typically a byproduct of DoS attacks on the network, refers to the delayed transmissions between controllers and the plant due to the unavailability of communication devices, communication paths, or local plant devices. Long et al. (Long, Wu, & Hung, 2005) numerically evaluated the impact of signal delays on a control performance of a proportional integral controller and a second-order plant. Two DoS attacks are modeled: one is the attack on a local controller to cause a large number of packet losses, and the other is a remote attack through Internet on a service-provider-edge router to cause a long delay jitter. The authors used a lumped queue to model the end-to-end packet transmission between a plant and a controller. The attack is injected as a packet traffic flow at different nodes of the network, and the signal delay is evaluated in terms of impacts on the control performance. Soucek et al. (Soucek, Sauter, & Koller, 2003) evaluated the effect of a delay jitter at a fixed mean delay on the quality of control. Two sources of the jitter delay are identified: network traffic-induced and protocol-induced.
- *Signal Blocking* refers to a situation in which the downstream receiver cannot receive the assigned signals. It is also considered a consequence of DoS attacks in many publications. Huang et al. (Huang, et al., 2009) considered signal blocking as a consequence of DoS attacks launched on a network-based control system. Sridhar and Manimaran (Sridhar & Manimaran, 2010) also explored a DoS attack that blocks the actuators from receiving real-time control actions from the controller.

Table 4. The common KPIs used in the selected papers.

| Author & Year | KPI | Description |
|---|---|---|
| (Fu, O'Neill, Yang, et al., 2021, Fu, O'Neill, & Adetola, 2021) | 1. Quality of building service<br>2. Quality of grid service | Energy Usage [kWh]; Peak Power Demand [kW]; Thermal Discomfort [Kh]; Demand Flexibility Indicator [kW] includes Upward Flexibility and Downward Flexibility |
| (Paridari, et al., 2016) | Financial impact | Energy Cost [EUR]; Degradation of Energy [%] |
| (Jacobsson, Boldt, & Carlsson, 2016) | Risk Values: Low, Medium, High | The risk values were calculated by multiplying the mean probability and the consequence values of identified risks based on the Information Security Risk Analysis (ISRA) questionnaire from two collaborative workshop sessions including security experts. Domain experts, and system developers. |
| (Bengea, et al., 2015) | Quality of building service | Energy consumption [kWh]; Peak Power Demand [kW]; Comfort violation [Kh] |

A few impact assessment frameworks have been developed for BASs. Kotenko and Chechulin (Kotenko & Chechulin, 2013) proposed a cyber attack modeling and impact assessment framework containing five main groups of security and impact assessment metrics. The first group includes metrics that are connected with topology, criticality, and vulnerabilities of the analyzed system (hosts): the level of the host vulnerability which is defined on the base of the known vulnerabilities. The second group includes metrics characterizing the attack, for example, attack potentiality. The metrics of the third group characterize the malefactor's potential and are intended to define possibilities of the attack development. The metrics of the fourth group are response efficiency and response collateral damage. The last group includes integral spatial characteristics of the system security and a score of the system risk level. Jacobsson et al. (Jacobsson, et al., 2016) proposed a risk analysis procedure based on six attributes, identifier, vulnerability, threat, probability value, consequence value, and risk value. The authors identified 32 risks and classified 9 risks as low, 19 risks as moderate, and 4 risks as high. Table 4 summarizes the Key Performance Indexes (KPIs) used to quantify the impacts of threats on BASs. The KPIs can be categorized into four types: economic impact, quality of building service, quality of grid service, and risk level.

## 3.4 Cyber-attack Detection and Defense Approaches for BASs

Section 3.4 reviews the detection and defense approaches for BASs in terms of three levels (i.e., management level, automation level, and field level) depending on where the detection or defense approaches are implemented. Through the literature review, we found that the implementation locations often overlap at both the management and automation levels. Data used for detection or defense are often collected from the automation level while the implementation is in the management level for its computing power. Implementations on the routers or standalone detection/defense devices are considered as part of the joint level between the management level and automation level. Thus, we will discuss the automation and management levels together.

### 3.4.1 Detection approaches

Table 5 summarizes the detection approaches for BAS cyber-attacks in recent publications. 54% of studies (e.g., 14 papers) used simulation data, 15% of studies (e.g., 4 papers) used HIL data, and 38% of studies (e.g., 10 papers) used field data. Despite simulation data being predominated, real data are highly needed for developing and validating convincing detection algorithms. Considering the challenges that launching cyber-attacks in real buildings may be unacceptable for building owners, a hardware-in-the-loop testbed could be a more feasible and efficient way for cyber-attack studies (Li, et al., 2022). In general, the detection methodologies reviewed in this paper can be grouped into rule-based (65% of studies), data-driven (34% of studies), and visualization-based (7% of studies) at the management & automation levels.

- *Rule-based Detection* (also called specification-based or signature-based detection in some literature). Current countermeasures to address the attacks mainly rely on network traffic screening and regularization. Esquivel-Vargas et al. (Esquivel-Vargas, Caselli, & Peter, 2017) described a specification-based intrusion detection system (IDS). The IDS first extracts BACnet implementation details from documentation of certified in-field devices, called the protocol implementation conformance statement, and then compares the network traffic with these rules to detect cyber intrusions. Fauri et al. (Fauri, et al., 2018) proposed an IDS for the BAS that detects known and unknown attacks, as well as anomalous behavior. A BACnet parser is used to extract the relevant message fields from each message in order to create a white-box model of the nominal system behavior. A human domain expert manually refined a collection of known BACnet threats into attack patterns.

Once an attack is detected, the system generates enriched alerts that include semantic information helpful to the operators. Čeleda et al. (Čeleda, Krejčí, & Krmíček, 2012) demonstrated the advantages of using a flow-based monitoring system and an entropy-based detection approach to detect security threats in the BACnet network through three use-cases in the Masaryk University Campus BAS network.

- *Data-driven Detection*. Peacock (Peacock, 2019) adopted machine learning algorithms of Artificial Neural Networks (ANN) and Hidden Markov Models to detect known and unknown network attacks based on a range of real and simulated datasets. Legrand et al. (Legrand, Niepceron, Cournier, & Trannois, 2018) proposed an autoencoder neural network that is used to measure the distance between a set of input and output vectors, establishing a threshold for anomaly classification.
- *Visualization-based Detection*. Besides the rule-based and data-driven methods, Tonejc et al. (Tonejc, Kaur, Karsten, & Wendzel, 2015) presented a visualization-based method for identifying application layer anomalies in BACnet based on network message flows. The approach is implemented as a mobile-based application for displaying application data and as a tool to analyze the communication flows using directed graphs. Thus, anomaly detection mainly relies on the users' experience in the BAS field.

Through the literature search as described in Section 2, we only found one publication that implemented a device-level detection solution at the BAS field level. Jones et al. (Jones, Carter, & Thomas, 2018) proposed an automated device-level solution that utilized unsupervised ANN to monitor BACnet networks and deployed a single board computer that can intercept communications between BAS devices at the field level. When an attack is detected, malicious traffic is blocked until the affected node is restored to its normal working state. However, implementing such a field-level solution with extra board computers could be costly.

Table 5. Review results on cyber-attack detection in BASs at different levels.

| BAS Levels | Author & Year | Detection Type | Approach Type | Validation | Summarized Highlights |
|---|---|---|---|---|---|
| Automation & Management Levels | (Elnour, Meskin, Khan, & Jain, 2021) | Detection of man-in-the-middle (MITM) attacks, unauthorized control commands | Data-driven: 1. Principal Component Analysis 2. Convolutional Neural Network Auto-Encoder | Simulation | A semi-supervised, data-driven isolation forest-based attack detection approach for a multi-zone HVAC system was proposed in which the normal operation data were used to develop the detection model. |
| | (Haque, Rahman, Chen, & Kholidy, 2021) | Detection of injecting false sensor measurements | Rule-based: Satisfiability Modulo Theory (SMT)-based solver | Simulation with real-world datasets | A control-aware attack analysis framework using a SMT-based solver to disclose vulnerable sensor measurements. |
| | (Peacock, 2019) | Anomaly Detection of Flood, DoS Reconnaissance, Write, and | Data-driven & Rule-based: 1. Hidden Markov Models 2. ANN | Simulation & Field test | Artificial Neural Networks and Hidden Markov Models were explored and found capable of detecting |

| | | | | |
|---|---|---|---|---|
| | Spoofing attacks. | | | known network attacks. Further, Hidden Markov Models were also capable of detecting unknown network attacks in the generated datasets. |
| (Sheikh, Kamuni, Patil, Wagh, & Singh, 2019) | Detection of DoS and Replay attacks | Data-driven: Support Vector Machine | Simulation | A Machine Learning algorithm and a Boolean Identification Strategy are proposed to identify whether the BAS operation is normal or faulty or under attack. |
| (Zhang, Kodituwakku, Hines, & Coble, 2019) | Detection of MITM, DoS, data exfiltration, data tampering, and false data injection attacks | Data-driven: Four classical classification methods including k-nearest neighbor, decision tree, bootstrap aggregating (bagging), and random forest | Hardware-in-the-loop (HIL) | A three-layers cyber-attack detection system consists of 1) firewalls and data diodes, 2) classification models based on network traffic and system data, and 3) empirical models based on physical process data. |
| (Hachem, Chiprianov, Babar, Khalil, & Aniorte, 2020) | Detection of fake emergency attacks, DDoS attacks | Rule-based: An extension modeling language named SysML to capture BAS vulnerabilities; A security extension of Multi-Agent Systems (MAS) to predict cascading attacks | Field test | A Systems-of-Systems Security (SoSSec) method that comprises: (1) a modeling language (SoSSecML) for secure SoS modeling and (2) MAS for security analysis of SoS architectures in buildings. |
| (Novikova, Bestuzhev, & Kotenko, 2019) | Detection of fabricating HVAC sensors readings | Visualization-based: A multivariate data visualization algorithm named RadViz | Simulation with real-world datasets | A RadViz-based visualization-driven approach to detect suspicious deviations in the system's state. |
| (Pan, Hariri, & Pacheco, 2019) | Detection of Who is attack, Read-Property attack, Write-Property attack, | Rule-based: Context Aware Data structure | HIL | A context aware intrusion detection framework which can accurately detect and classify different types of BAS |
| | I-Am attack, BACnet Routing attack, False Alarm attack, Flooding attack, Malfunction, Reinitialize Device Attack | | | attacks and asset malfunctions. |
| (Belenko, Chernenko, Kalinin, & Krundyshev, 2018) | Intrusion Detection | Data-driven: Generative Adversarial Networks (GAN) | Simulation | Generative adversarial ANNs to detect security intrusions in large-scale networks of cyber devices. |
| (Fauri, et al., 2018) | Intrusion Detection of Snooping, Tampering, Spoofing, DDoS attacks | Rule-based: A BACnet parser using a white-box model | Simulation & Field test | An intrusion detection system of detecting known and unknown attacks as well as anomalous behaviors for BASs by leveraging protocol knowledge and specific BACnet semantics. |
| (Legrand, et al., 2018) | Anomalies detection of Peak/Point Anomalies, Contextual Anomalies, Collective Anomalies | Data-driven: Convolutional and Recurrent autoencoder neural networks for outlier detection | Field test | Two types of autoencoder neural networks are used to measure the distance between a set of input and output vectors, establishing a threshold for anomaly classification. |
| (Zheng & Reddy, 2017) | Detection of Abnormal network traffic, IP Spoofing and Data Injection, Session Hijacking, Reconnaissance Attack, DoS Attack, Safety-critical Attack | Rule-based: Flow-service models for time-driven traffic | Real-world BACnet traffic in BAS networks | An anomaly detector named THE-Driven for BACnet that is able to detect suspicious traffic in BAS networks considering three types of traffic: time-driven, human-driven, and event-driven traffic. |
| (Esquivel-Vargas, et al., 2017) | Intrusion Detection of Backdoor, Active Device Fingerprinting, DoS attacks | Rule-based: Specification-based intrusion detection at the network level, specifications are individually tailored for each | Field test | A parsing method is developed for BACnet protocol to detect specification-based intrusion based on two-month real BACnet traffic. |
| | | device in the network | | |
| (Harirchi, Yong, Jacobsen, & Ozay, 2017) | Fraud Detection of sensor data injection attack | Rule-based: Active model discrimination | Simulation | An active model enables a system operator to detect and uniquely identify potential faults or attacks in a potential utility bill fraud scenario. |
| (Pan, et al., 2016) | Intrusion Detection of Who-Is/Who-Has attack, Write/Write-Multiple attack, Flooding Conformed Service attack, I-Am attack | Rule-based: A data mining algorithm called Decision Tables is applied to generate the classification rules to dynamically classify target assets and attack mechanisms | Field test | An anomaly-based intrusion detection system that monitors BACnet traffic utilizing two novel data structures: Protocol Context Aware and Sensor-DNA. |
| (Paridari, et al., 2016) | Intrusion Detection of MITM attacks | Data-driven & Rule-based: 1. Reduced order model, Threshold-based outlier detection 2. Machine-learning outlier detection | Simulation | A physical approach to detect anomalies and outliers using the measurement data. |
| (Caselli, Zambon, Amann, Sommer, & Kargl, 2016) | Intrusion detection of Process Control Subverting, Snooping, and DoS attacks | Rule-based: Specification-mining approach to generate specification rules | Field test | A specification-based network intrusion detection system for BACnet-based building automation systems that can used to demonstrate a specification mining approach for network security monitoring. |
| (Xu, Wang, & Jia, 2016) | Detection of Abnormal network traffic | Rule-based: Counting Bloom Filter and Compressed Bloom Filter | Real-word network traffic from distributed home networks | A bloom-filter based analytics framework to capture persistent threats towards the same home routers and to identify correlated attacks towards distributed home networks. |
| (Al Baalbaki, Pacheco, Tunc, Hariri, | Detection of DoS, Delay, Flooding, Network Knockdown, | Rule-based: Feature Extraction and Rule Generation based | Simulation | An anomaly behavior analysis system for ZigBee protocol to be used to detect known and unknown ZigBee attacks. |
| & Al-Nashif, 2015) | Jamming and Pulse DoS attacks | classification model | | |
| (Kaur, et al., 2015) | Intrusion Detection -DoS, Flooding Attack, Smurf Attack, Traffic Redirection Attack | Rule-based: A Snort-Based BACnet Normalizer extension capable of normalizing BACnet/IP traffic based on a configuration file. | Simulation | A snort-based traffic normalization method for improving application reliability and security of BACnet. |
| (Liu, et al., 2015) | Detection of Abnormal field data | Data-driven:<br>1. Short-term detection: support vector regression<br>2. Long-term detection: partially observable Markov decision process | Simulation | A smart home energy pricing cyber-attack detection framework which integrates the net metering technology with short/long term detection. |
| (Tonejc, et al., 2015) | Anomalies Detection based on the users' experience in the BAS field | Visualization-based: Visualizing network message flows to facilitate humans in building-security decision-making | HIL | A visualization method for identifying application layer anomalies in BACnet based on network message flows. |
| (Pan, Hariri, & Al-Nashif, 2014) | Anomaly-based intrusion detection of Reconnaissance Attack, Device Access Attack, DoS Attack | Rule-based: Attack classification based on a set of rules that characterizes the BACnet behavior | Simulation | An intrusion detection framework consists of four modules: Monitoring module, Training module, Attack Classification module, and Action Handler module. |
| (Čeleda, et al., 2012) | Intrusion Detection of BACnet router spoofing attack, DoS attack, Write attack | Rule-based: Entropy-based detection approach | Field test | An entropy-based approach of detecting network anomalies compared with simple volume based approaches |

| | | | | | |
|---|---|---|---|---|---|
| | (Wendzel, Kahler, & Rist, 2012) | Intrusion Detection and Prevention | Rule-based: A prototype based on the BACnet firewall router to implement multi-level security in BACnet environments | Simulation | A BACnet Firewall Router of detecting and mitigating covert storage and covert timing channel attacks. |
| Field Level | (Jones, et al., 2018) | Intrusion Detection, Security Monitoring | Data-driven: unsupervised Artificial Neural Networks (ANN) | HIL | An automated device-level solution to secure BACnet networks. |

Note: The detection approaches for the management level and automation level are reviewed in one category since most of them rely on resources (data, software or hardware) from both levels.

### 3.4.2 Defense approaches

Table 6 summarizes the defense approaches against BAS cyber-attacks in terms of three levels. 38% of studies (e.g., 6 papers) used simulation data, 25% of studies (e.g., 4 papers) proposed conceptual approaches without BAS data, and 43% of studies (e.g., 7 papers) used field data. Most papers used field data, which are more convincing, to develop and validate their proposed defense algorithms. In general, the defense approaches at the field level mainly focus on privacy protection and device verification. The defense approaches at the management & automation levels can be summarized into BAS protocol hardening, network firewall, and traffic normalization. Protocol hardening is to add security features to protocols. A network firewall is to block illegal traffic. Traffic normalization is to correct traffic based on normalization rules extracted from protocol standards and implementation specifications.

- *BAS protocol hardening*. The attack surface of a system is the set of ways in which an adversary can enter the system and potentially cause damage (Manadhata & Wing, 2010). The smaller the attack surface, the easier it is to protect. As BASs get integrated into existing IP-based networks or even communicate directly over the internet, the attack surface of BASs has increased dramatically and thus BASs require a solid security architecture. According to this context, Judmayer et al. (Judmayer, Krammer, & Kastner, 2014) reviewed and compared two security extensions, KNXnet/IP Secure (Gützkow, 2022) published by the KNX association and the generic security concept proposed by Granzer et al. (Granzer, Lechner, Praus, & Kastner, 2009). Yang et al. (Yang, et al., 2022) proposed a module to prevent attackers from performing DDoS attacks and a transport layer security protocol with an encrypted token identity authentication module to ensure internet security in the energy management system. Shang et al. (Shang, Ding, Marianantoni, Burke, & Zhang, 2014) proposed a data-centric, encryption-based, and non-interactive approach enabled by the named data networking architecture to secure BAS network communications. ASHRAE SSPC-135 IT Working Group (IT-WG) has developed a new proposal centered on secure communications exclusively using accepted IT best practices called "BACnet Secure Connect (BACnet/SC)" (Fisher, Isler, & Osborne, 2019).

BACnet/SC eliminates the need for static IP addresses and network broadcasts, and provides secure message transport using the standard IP application protocol.

- *BAS network firewall and traffic normalization*. ur Rehman et al. (ur Rehman & Gruhn, 2018) implemented a secure firewall between the LAN and the Internet Service Provider (ISP), for protecting IoT environments. The firewall is able to defend against malicious programs, unauthorized access, and DoS attacks. Fovino et al. (Fovino, Coletta, Carcano, & Masera, 2011) adopted a ModBus firewall, which sits between the master and slave devices on a network to monitor the critical state of the system. Alerts are generated based on legitimate commands by monitoring the evolution of the state of the protected system and analyzing the command packets between master and slaves of a SCADA architecture. The ModBus firewall could block Unauthorized Command Execution, DoS, MITM, and Replay attacks. Wang et al. (Wang, et al., 2015) proposed a security/safety modeling framework using proxy-based policy enforcement and formal verification, which enables blocking attacks made towards embedded BAS controllers. Kaur et al. (Kaur, et al., 2015) proposed a Snort-based BACnet normalizer which enforces the BACnet rules in the network traffic captured by the Snort agent.
- *Field-level securing solutions*. Occupancy sensors collect occupancy data to enable intelligent HVAC controls adapted to occupancy variations. However, an adversary with malicious intent could exploit occupancy data in combination with auxiliary information to infer privacy details about indoor locations of building users. To protect individual location information from being inferred from the occupancy data, Jia et al. (Jia, Dong, Sastry, & Sapnos, 2017) proposed a privacy-enhanced architecture that distorts the occupancy data to hide individual occupant location information while maintaining HVAC performance. Wireless Sensor Networks (WSNs) are commonly utilized to monitor wireless field devices in critical infrastructure applications such as hospital buildings, where WSNs can track expensive medical equipment and patient stay and continuously monitor patient vital signs. However, the nature of the wireless broadcast medium enables potential attackers to conduct active and passive attacks. Dubendorfer et al. (Dubendorfer, Ramsey, & Temple, 2013) introduced radio frequency fingerprinting techniques to detect and reject unauthorized rogue devices in WSNs. Formal verification is commonly used to secure filed devices, especially embedded devices. Antonini et al. (Antonini, Barenghi, et al., 2014) highlighted a field-level formal code verification approach to provide safety and security for Programmable Logic Controller code in SCADA and BASs.

Table 6. Review results of cyber-attack defense in BASs at different levels.

| BAS Levels | Author & Year | Defense Type | Approach Type | Validation | Summarized Highlights |
|---|---|---|---|---|---|
| Automation & Management Levels | (Yang, et al., 2022) | Defense of DDoS, MITM, replay, and impersonation attacks | Trusted Encrypted Validator Module based on Token Authentication | Simulation and Field test | An encrypted token identity authentication module enables preventing attackers from performing DDoS attacks on the energy management system by encrypting, decoding, and verifying the device's legality. |

| | | | | |
|---|---|---|---|---|
| (Yahyazadeh, Podder, Hoque, & Chowdhury, 2019) | Blocking undesired implicit interplay, explicit interplay, sneaky commands, contextually benign commands | A platform-agnostic formal specification language is used to encode the users' expectation of the building automation behavior, thus defining a set of policies that are later used to verify actions and validate app behavior. | Field test | A framework named Expat aims at protecting smart-home platforms from malicious automation apps. |
| (ur Rehman & Gruhn, 2018) | Defense against malicious programs, unauthorized access, DoS attacks | A sicher firewall detects and generates warnings to users and invokes mitigation strategies against particular security breaches | Concept | A sicher firewall acts like a filter between the net/LAN and the Internet Service Provider (ISP) for protecting smart home and IoT environments. |
| (Airehrour, Gutierrez, & Ray, 2016) | Defense against blackhole routing attacks | A trust-based mechanism | Simulation | A trust-based routing protocol provides a feedback-aware security system for IoT networks. |
| (Wang, et al., 2015) | Hardware/Software Defense of false data injection attack, resource consumption attack, deception attack, replay attack, and DoS attack | A microkernel structure including a trusted platform module, proxy-based policy enforcement, and formal verification | Field test | A security/safety modeling framework enables blocking attacks made towards embedded BAS controllers by adopting a microkernel-based architecture. |
| (Sparrow, Adekunle, Berry, & Farnish, 2015) | Cryptography-based Defense | Mathematical models | Simulation | Two security mechanisms with a focus on Authenticated Encryption with Associated Data can secure wireless sensor multi-hop networks. |

| | | | | | |
|---|---|---|---|---|---|
| | (Judmayer, et al., 2014) | Protocol-specific Defense | Symmetric cryptography mechanisms using the Advanced Encryption Standard with 128-bit as a block cipher | Concept | Two security extensions for IP-based KNX networks. |
| | (Shang, et al., 2014) | Identity-based access control to enforce trust relationships and uses encryption to protect against unauthorized reads | A hierarchical namespace for data, encryption keys, and access control lists | Field test | A data-centric BMS design that uses information-centric networking architecture designs to secure network communications. |
| | (Hager, Schellenberg, Seitz, Mann, & Schorcht, 2012) | Cryptography-based Defense | Hash algorithms, authentication methods, and a role-based access system | Simulation | A complete communication architecture of securing smart homes to authenticate each participant and restrict access to all the data and functions of the system |
| | (Fovino, et al., 2011) | Defense of Unauthorized Command Execution, DOS, MITM attacks, Replay Attacks | Critical state based filtering method | Field test | A network filtering approach for the detection and mitigation of a particular class of cyberattacks against industrial installations. |
| | (Muraleedharan & Osadciw, 2006) | Defense against DoS attacks | Swarm intelligence based approach | Simulation | To prevent DoS attacks from wireless sensor networks, Swarm intelligence is applied to detect the possible routing and the best routing performances. |
| Field Level | (Jia, et al., 2017) | Protect individual location information of occupancy-based HVAC controllers | Optimization-based method by formulating the privacy-utility trade-off problem that minimizes the privacy loss | Real-world occupancy data and Simulated building dynamics | A privacy-enhanced framework uses occupancy-based HVAC control as the control objective and the location traces of individual occupants |

| | | | | |
|---|---|---|---|---|
| | | subject to a pre-specified controller performance constraint | | as the private variables. |
| (Antonini, Barenghi, et al., 2014) | Formal Verification based solutions to protect field devices | Formal verification with safety constraints | Concept | A survey of formal verification solutions to secure devices on the SCADA and BASs. |
| (Kanuparthi, Karri, & Addepalli, 2013) | Secure IoT in terms of four key challenges, 1) data provenance and integrity, 2) identity management, 3) trust management, and 4) privacy | Embedded and hardware security approaches: Physical unclonable function, Hardware performance counters, and Lightweight encryption algorithms | Concept | Physical Unclonable Function technology is used for data provenance and integrity, and identity management. Hardware performance counters are used for trust management. Lightweight cryptography is used to provide privacy. |
| (Dubendorfer, et al., 2013) | Defense of unauthorized rogue devices in ZigBee network | Radio Frequency (RF) fingerprinting techniques | Field test | An ID verification method with dimensionally-efficient RF fingerprints can detect and reject unauthorized rogue devices. |
| (Bordencea, Valean, Folea, & Dobircau, 2011) | Defense of failure sensors | An adaptive and fault-tolerant system using Paxos protocol to allocate the sensors to Access Points (APs) under churn. When an AP fails, its role will be taken by another AP. | Field test | A software agent based system providing adaptation and fault tolerance allows a system to continue to function in presence of access point failure or defective sensors. |

Note: The defense approaches for the management level and automation level are reviewed in one category since most of them rely on resources (data, software or hardware) from both levels.

### 3.4.3 BAS security framework and guideline

This section highlights a practical framework and a guideline applicable to BAS security from the available literature.

The NIST developed a cybersecurity framework for critical infrastructure to identify, assess, and manage cyber risks (Barrett, 2018). The U.S. Department of Energy's Pacific Northwest National Laboratory developed the Buildings Cybersecurity Framework (BCF) (Cybersecurity, 2018, Mylrea, Gourisetti, & Nicholls, 2017) to secure BASs based on five core elements defined by the NIST cybersecurity framework: Identity, Protect, Detect, Respond, and Recover, as shown in Figure 6. The goal of the *Identify* function is to identify cyber risks and vulnerabilities and then develop the organizational capacity to manage cybersecurity risk to systems, assets, data, and capabilities. The goal of the *Protect* function is to protect assets by introducing building operators to cyber protection techniques. The goal of the *Detect* function is to highlight techniques that enable the detection of malicious cyber activity. The goal of the *Respond* function is to respond to a cyber-attack by developing and implementing the appropriate processes to respond to a cybersecurity incident effectively. The goal of the *Recover* function is to recover and return services to normal operation and reduce the impact of a cybersecurity event.

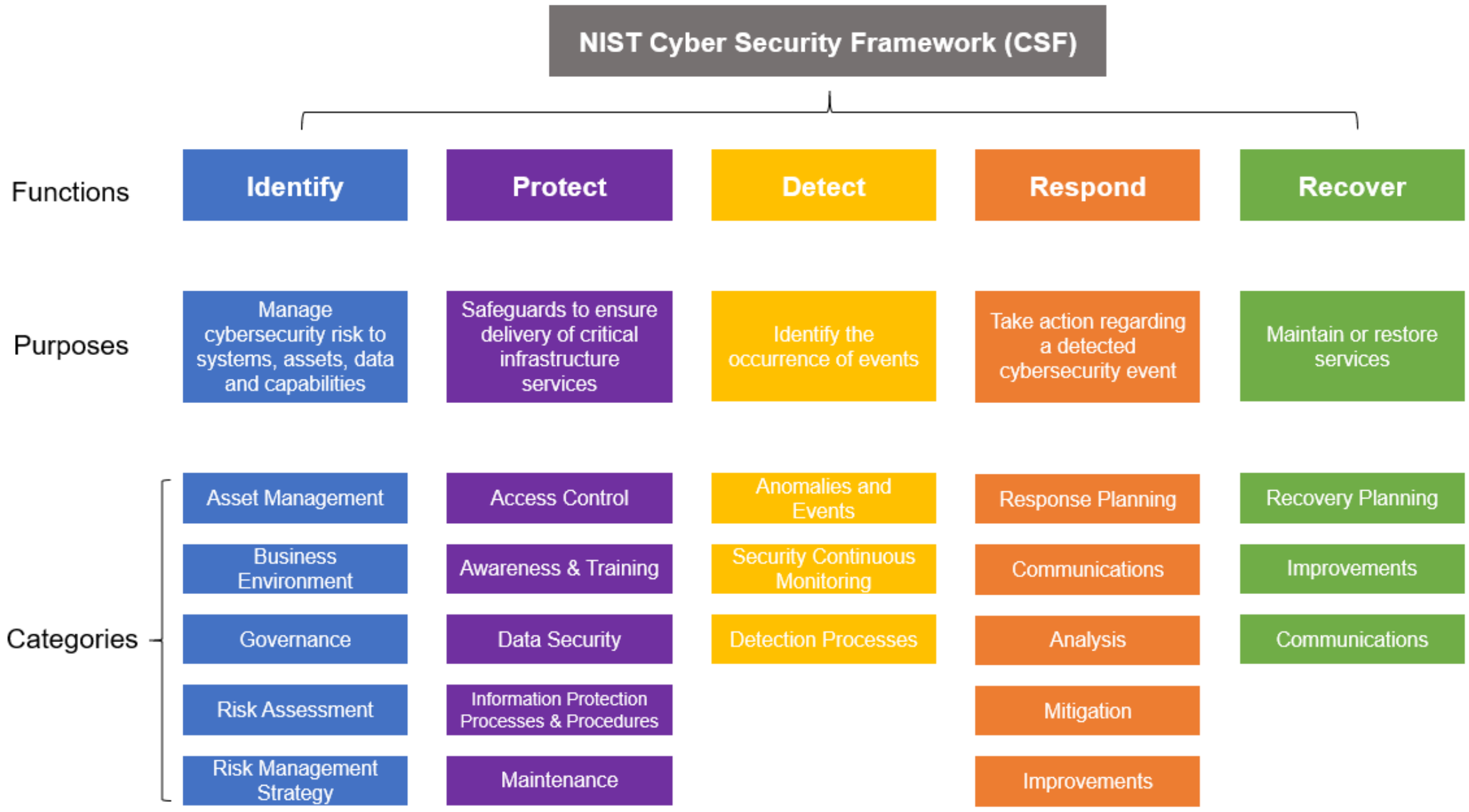

Figure 6. Summary of the NIST cybersecurity framework.

The MITRE Adversarial Tactics, Techniques, and Common Knowledge (ATT&CK) (Strom, et al., 2018) is a guideline for classifying, describing, and tackling cyberattacks and intrusions for industrial control systems, which is also applicable to BASs. To address the lack of attack-defense mapped frameworks, Kwon et al. (Kwon, Ashley, Castleberry, Mckenzie, & Gourisetti, 2020) presented a tool called the "Cyber Threat Dictionary (CTD)" to provide immediate solutions to practitioners by mapping ATT&CK Matrix to the NIST cybersecurity framework. CTD can be used in both reactive and proactive ways. For reactive usage, cybersecurity practitioners can identify corresponding actions once an attack is detected. For proactive usage, practitioners can utilize CTD to identify how controls will defend against possible attacks and identify gaps before controls are exploited.

## 3.5 Cyber resilient control

While cyber detection and defense techniques can help reduce cyber-attack risk, immunity to known and unforeseen malicious activities is not guaranteed. Cybersecurity and cyber resilience strategies are most effective when combined. A resilient cyber-physical system (CPS) is one that maintains state awareness and an acceptable level of operational normalcy in response to disturbances, including threats of an unexpected and malicious nature (Rieger, Gertman, & McQueen, 2009). A cyber resilient control strategy can help mitigate the impacts of successful attacks on BASs. However, through the literature review, we found only a few publications on cyber resilient control for buildings. Implementing and evaluating cyber resilient control strategies for buildings are limited in practice. It's worth mentioning that the existing advanced control technologies can significantly improve the BAS cyber-resiliency when informed by cyber-detection outcomes.

As detailed in the following sections, the different methodologies to achieve resilient control can be broadly classified as passive or active. Passive methods restrict their attention to threats that can be characterized and modeled offline. The controls are designed to enable the closed-loop system to tolerate anticipated abnormalities and rely only on sensor feedback to attenuate the impact of a threat. On the other hand, active methods react to threats by taking advantage of real-time information. Real-time situation awareness is combined with control methods to handle system abnormalities or disruptions. As a result, they are reconfigurable and more effective at mitigating unforeseen events. A representative schematic diagram of the resilient control methods is depicted in Figure 7.

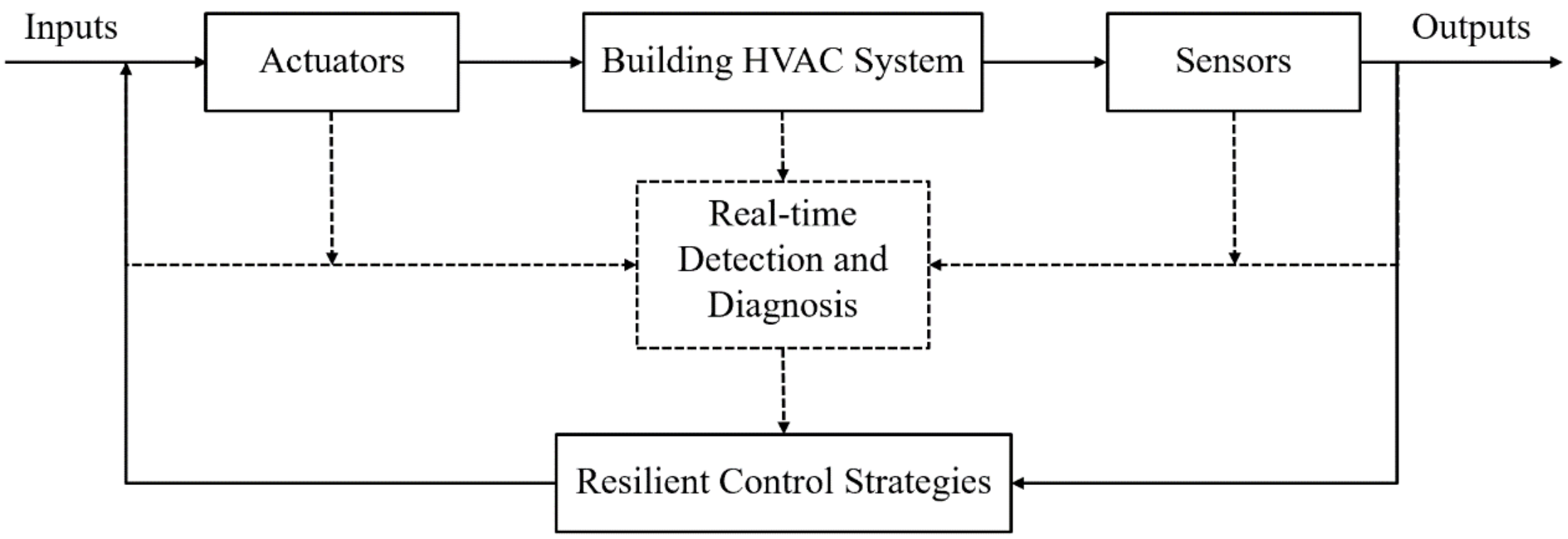

Figure 7. Schematic diagram of resilient control (adapted from (Gao & Liu, 2021)).

### 3.5.1 Passive resilient control – Fixed controller

By regarding and modeling the abnormal signature of cyber-attacks on building systems as disturbances or uncertainties, robust controllers can be designed to mitigate the consequences of abnormalities and provide passive resilient control (Zhang & Jiang, 2008). Weerakkody et al. (Weerakkody, Ozel, Mo, & Sinopoli, 2019) proposed a robust design of distributed control system to balance the costs of sensing and communication with the need for security. Huang and Wang (Huang & Wang, 2008) presented a two-loop robust Model Predictive Control (MPC) framework for HVAC temperature control. The inner-loop controller ensures robust stability of the local loops using a classical controller while the outer-loop controller improves the overall control performance based on the predicted system information and by accounting for the uncertainties and constraints of the HVAC system. Wang and Xu (Wang & Xu, 2002) presented a robust control strategy to address instability issues when transitioning between different control modes in building HVAC applications. Other works, such as (Bengea, et al., 2015, Homod, 2014, Lebreton, Damour, Benne, Grondin-Perez, & Chabriat, 2016) have studied passive control strategies that can tolerate physical faults and maintain normal or critical building operations. While these methods did not target cyber-attacks, they could be effective solutions for mitigating the impact of cyber threats with similar signatures and effects on the building systems. In general, passive resilient control methods can handle a broad range of system abnormalities, but they may be overly conservative, resulting in poor performance under threat-free operations (Teixeira, Kupzog, Sandberg, & Johansson, 2015).

### 3.5.2 Active resilient control – Reconfigurable controller

Integrating attack detection mechanisms and reconfigurable control methods is a possible approach to ensure system resilience against cyber-attacks. Chen et al. (Chen & Shi, 2021) proposed a Stochastic MPC (SMPC)-based resilient secure control framework, which consists of an attack detector, a resilient estimator, and a resilient SMPC controller. The attack detector serves as the decision-making module for triggering the resilient control. If a DoS or deception attack is detected, the resilient estimator estimates the unobserved state based on tampered states, and the resilient SMPC controller will be selected to compute the control actions; otherwise, the SMPC controller will work in the normal mode. Sun et al. (Sun, Zhang, & Shi, 2019) designed a resilient MPC framework for cyber-physical systems (CPSs) under DoS attacks, where the CPS was modeled as a linear time-invariant system. A conventional dual-mode MPC strategy was adapted to handle the attack and the physical system constraints simultaneously. An optimization-based control was used to steer the system state into a predefined terminal set, and then a state-feedback control law was applied to maintain stability after the state entered this set. Considering DoS attacks corrupt the communication channel between the controller and the actuator, the maximum tolerable duration of the attacks under which the closed-loop system remains stable was established.

Estimation-based resilient control methods have been proposed for BAS resiliency against cyber-attacks. Paridari et al. (Paridari, et al., 2016) presented a resilient hierarchical control framework for addressing adversarial actions on sensor measurements. The control policy used estimated values, rather than the corrupted measurement, to drive the control decisions. Paridari et al. (Paridari, et al., 2017) further proposed a data-driven anomaly detection method and a control reconfiguration strategy to maintain the system stability and performance under man-in-the-middle sensor attacks. The resilience control policy is based on corrected measurement signals estimated from virtual sensors. Since the virtual sensor adaptation and controller reconfiguration algorithms are implemented at the supervisory layer, the system does not require major modification to the local controllers. Xu et al. (Xu, Fu, Wang, O'Neill, & Zhu, 2021) presented a machine learning-based framework for sensor fault detection and mitigation. The proposed sensor fault-tolerant framework includes three neural network-based components for generating temperature predictions in different ways with the consideration of possible sensor faults, selecting one of the predictions based on the assessment of their accuracy, and applying reinforcement learning for HVAC control based on the selected prediction, respectively.

State resetting and fallback mechanisms can support system resiliency. Feng et al. (Feng & Tesi, 2017) investigated the problem of designing DoS-resilient control architectures for networked systems. It was shown that the use of dynamical observers with a measurements-triggered state resetting mechanism can enable the system to tolerate a general class of DoS attacks. The authors adopted dynamic controllers equipped with prediction and state resetting capabilities. The prediction capability compensates for the lack of data during DoS periods, while the state resetting provides fast state reconstruction.

In general, the rationale behind the active approaches is to adapt or reconfigure the control system only when an attack has been detected and diagnosed while avoiding a complete redesign of the control algorithms to ensure a good performance under nominal conditions. The overall objective of control reconfiguration is to minimize the loss in performance inflicted by attacks while maintaining an acceptable level of operational normalcy. Although some of the aforementioned resilient control strategies (Chen & Shi, 2021, Feng & Tesi, 2017, Paridari, et al., 2017, Sun, et al., 2019, Weerakkody, et al., 2019) are not specifically designed for BASs, they can be extended to empower BASs with cyber-attack-immune capabilities.

## 4. Open Challenges and Future Opportunities

Based on our comprehensive literature review, we identified a set of open challenges and future opportunities that we believe deserve further attention from the research community on BAS security in commercial buildings.

***Challenges***

1) Handling the growing complexity and different protocols of BASs as more sensors and actuators are being included (Ciholas, et al., 2019) in modern intelligent buildings in the era of IoT.
2) Conducting realistic experiments and field demonstration to evaluate the cyber-secure strategies. It is difficult to convince building owners and building facility teams to lend their buildings for cyber-attack testing. At the same time, the scalability and interoperability of the current cyber-security solutions is limited considering a variety of communication protocols and BAS with proprietary hardware and software.
3) Advancing the convergence of IT and OT technologies of BAS. Existing efforts, such as adding encryption (common IT practice) to BACnet protocol (common OT protocol), have enhanced BAS cyber-security. More efforts, such as BAS-specific network intrusion detection/prevention and malware detection, are still needed.
4) Persuading building owners to update their obsoleted BAS. Most BASs in existing buildings are designed to be used for decades with little consideration of cyber security. Hardware such as legacy devices may have difficulty upgrading with cyber-secure technologies due to limited memory and processing power. The investment cost of upgrading and implementation also plays a vital role in the decision-making stage, influencing the motivation of the building owners.
5) Dealing with human factors in cyber-physical security studies. People-related issues require more attention, given the lack of security awareness of vendors, customers, and operators.
6) Leveraging advanced machine learning techniques (e.g., deep reinforcement learning) for data-driven intrusion detection and control in BAS in a trustworthy manner. The learning techniques that are effective in other domains often face significant challenges in practical operation of BAS, e.g., long training time and lack of data labels, high degree of data noise due to sensor faults and possible attacks, and lack of assurance in system robustness and reliability.

7) Lack of holistic cyber-physical modeling and analysis framework for investigating the effects of cyber-originated abnormalities on the operation of building HVAC systems.
8) Lack of quantifiable metrics and methods for assessing the resilience of BAS in terms of its ability to withstand and recover from successful cyber-attacks.

***Opportunities***

1) Developing real testbeds and generating realistic datasets. Launching cyber-attacks in a real building may not be acceptable for building owners. A hardware-in-the-loop testbed is a more feasible and efficient way for cyber-attack studies.
2) Developing cyber analytics solutions that can minimize the frequency of detection false alarms and accurately diagnose and localize cyber-attacks. Preventative strategies are needed as early alarms to catch cyber-attacks before they happen on BASs. Solutions that can differentiate cyber-attacks from physical faults are also needed to assure targeted response and fast recovery from the effects of adversarial events.
3) Conducting impact analysis to select a set of critical signals or devices for enhanced cyber hardening, thus achieving the most effective defense-in-depth cyber protection.
4) Developing resilient strategies that can handle multiple simultaneous cyber-attacks and physical faults. Most studies focused on only one type of event at a time. However, multiple cyber-attacks and physical faults can occur simultaneously. Therefore, an attractive future direction is developing a flexible detection/defense/control solution to tackle diverse and concurrent cyber threats and faults.
5) Developing machine learning techniques that are data efficient, fault tolerant, and robust in uncertain environment. One possible direction is to explore hybrid approaches that combine neural network-based methods (e.g., deep reinforcement learning) with physical models and rules developed by domain experts.
6) Developing building-specific cyber resilient control strategies. Few publications apply resilient control specifically to BASs autonomous and adaptive cyber response. Existing advanced control technologies have been proven to be successful at mitigating cyber-attack impacts in industrial control systems. This provides a practical opportunity to enhance the cyber-resiliency of buildings, especially critical infrastructures such as data centers, hospitals, and military bases.

## 5. Conclusions

This paper presented a comprehensive review integrating BAS vulnerabilities, potential threats with impact assessment, cyber-attack detection & defense approaches, and cyber resilient control strategies. In this paper, the hardware and software architecture of BASs are grouped into three levels: management, automation, and field. Then the general BASs vulnerabilities and protocol-specific vulnerabilities for the four dominant BAS protocols (i.e., BACnet, KNX, LonWorks, and Modbus) are reviewed, followed by the discussion on potential threat scenarios and impact assessment. Four attack targets (i.e., management devices running on IT network, interface from IT to OT network, interface from OT to IT network, and field devices) and seven potential attack scenarios are identified. The impact of cyber-attacks on BASs is summarized as signal corruption, signal delaying, and signal blocking. The typical cyber-attack detection and defense approaches are identified at the management & automation levels and the field level. Cyber resilient control strategies for BASs under attack are categorized into passive and active resilient control schemes. Finally, insights on open challenges and future opportunities are provided. With a comprehensive review, this paper provides critical information that could help transfer cyber-physical security technologies to the building industry.

## Acknowledgment

The research reported in this paper was partially supported by the Building Technologies Office at the U.S. Department of Energy through the Emerging Technologies program under award number DE-EE0009150.